# Frequency Domain Simulations of Charge-Density-Wave Strains: Comparison with Electro-Optic Measurements

L. Ladino and J.W. Brill
Department of Physics and Astronomy
University of Kentucky, Lexington, KY 40506-0055

**Abstract**
We have studied changes in charge-density-wave strain under application of square-wave currents of variable amplitude and frequency by numerically solving the phase-slip augmented diffusion model introduced by Adelman *et al* (Phys. Rev. B 53, 1833 (1996)). The frequency dependence of the strain, at each position and amplitude, was fit to a modified harmonic oscillator expression, and the position and current dependence of the fitting parameters determined. In particular, the delay time (1/resonant frequency) vanishes adjacent to the contact and grows with distance from the contact, and both the delay time and relaxation time decrease rapidly with increasing current (and phase-slip rate), as experimentally observed in the electro-optic response of blue bronze. We have also found that pinning the phase at the contacts causes more rapid changes in strain between the contacts than allowing the phase to flow outside the contacts.

Sliding charge-density waves (CDWs) in quasi-one dimensional conductors exhibit some of the most unusual phenomena in condensed matter physics.[1,2] A CDW emerges upon a decrease in temperature as a result of the Peierls transition, in which an energy gap forms on part (e.g. NbSe$_3$) or all (e.g. blue bronze K$_{0.3}$MoO$_3$) of the Fermi surface.[1] In the CDW state, a periodic lattice distortion is accompanied by a modulation in the electron density:

$$n = n_c + n_1 \cos[Qx + \chi(x,t)] , \quad (1)$$

where Q is the CDW wave vector, n$_c$ the density of electrons condensed in the CDW state, and n$_1$ and $\chi$ are the amplitude and phase of the CDW. In a perfect crystal, a CDW with Q incommensurate to the reciprocal lattice would have no preferred phase so that the CDW could slide freely through the crystal, resulting in collective current proportional to $\partial\chi/\partial t$. However, in the presence of defects, the CDW lowers its energy by elastically deforming, pinning the phase. When an electric field $E$ greater than a threshold field is applied, the CDW strains, i.e. becomes compressed near one contact and rarefied near the other, and slides between the contacts.[1,3-5]

There has been considerable interest in the question of how electrons enter and leave the CDW at current contacts.[3-5] CDW current flow requires a mechanism for adding or removing CDW wave fronts at the contacts to achieve conversion between collective and single-particle current. This mechanism is provided by phase-slip (creation and motion of CDW dislocations), in which rapid localized $2\pi$ changes in phase (corresponding to two electrons/conducting chain) are added or subtracted.[6] To account for these, Adelman et al[3] introduced a smoothly varying "renumbered" phase,

$$\phi = \chi - \int\int r_{ps}(x',t') \, dx' \, dt' \quad (2)$$

where $r_{ps}(x) = - Q/(en_c) \, \partial j_c/\partial x$ is the local rate of phase-slip and j$_c$ is the collective current density. Phase-slip, in turn, is driven by strain of the CDW phase, $\varepsilon \equiv Q^{-1} \partial\phi/\partial x$, which reduces the energy barrier to the formation of phase dislocation loops.[3]

Adelman et al[3] used multi-contact transport measurements to determine the position and time dependence of the strain in NbSe$_3$ after a current reversal, and simulated their results using a one-dimensional diffusion-like model for the phase:

$$\partial\phi/\partial t = Q/[en_c(\rho_{CDW} + \rho_0)] \, [\rho_0 \, j_{total} - E_P(j_c) + QK/(en_c) \, \partial^2\phi/\partial x^2] - \int r_{ps}(x',t) \, dx' \quad (3)$$

where $\rho_{CDW}$ and $\rho_0$ are the high field CDW and single particle resistivities, $K$ the CDW elastic constant, and j$_{total}$ is the total current density, i.e. j$_{total}$ = j$_c$ + j$_s$ where j$_s$ is the single particle current density. In this simplified model, the randomly distributed impurities have been replaced by a phenomenological pinning field, which may depend on CDW current, as discussed below. For their simulations, they approximated the strain dependence of the phase-slip rate by:[7]

$$r_{ps}(x) = - sgn[\varepsilon(x)] \, r_0 \exp[-\varepsilon_B/|\varepsilon(x)|] , \quad (4)$$

where $r_0$ and $\varepsilon_B$ are proportional to the attempt rate and barrier height for dislocation nucleation.[3] This phenomenological model captures the results of more microscopic calculations which considered the details of charge conversion and dislocation motion and pinning,[4] and reproduced the observed spatial/temporal phase variations of their experiments. In particular, Adelman *et al* found that the strain varied linearly with position in the center of the sample, with extra strains near (~ 100 μm) the contacts;[3] the strains near the center can be viewed as a finite-size effect and are inversely proportional to the length of the sample.[4]

We have observed similar strain profiles in blue bronze using an infrared electro-optic technique, in which the electro-optic response was assumed to be proportional to the strain of the CDW phase.[5,8,9] For these measurements, symmetric, bipolar square-wave voltages of variable frequency (ω) were applied to the sample, and the electro-optic response measured as functions of ω, voltage, and position. To enhance our understanding of these frequency dependent measurements and the evolution of parameters that govern CDW strain dynamics, we have numerically solved Eqtn. (3) for applied square waves. In addition, we analyze the effects of different boundary conditions on the dynamics of the strain. The results of these simulations are the subject of this paper.

For a square-wave current of a given frequency (ω) and amplitude, Eqtn. (3) was solved starting from $\phi = 0$ everywhere and repeating current reversals until the phase variation becomes periodic in time, $\phi(x,t) = -\phi(x,t+\pi/\omega)$. For each time step ($\Delta t \ll 1/\omega$), Eqtn. (3) was first solved without the phase slip term, which was then added. Finally, to mimic frequency-dependent lock-in amplifier measurements of electro-optic signals, we calculated, for each current and position, the fundamental term (in-phase and in quadrature with the applied square wave) of the Fourier series expansion of the time dependence, $\varepsilon_\omega$. For concreteness, Equation (3) was solved using the same parameters as used by Adelman *et al* for NbSe$_3$:[3] L = 670 μm (= distance between current contacts), $\rho_0$ = 8.8 x 10$^{-7}$ Ω·m, $\rho_{CDW}$ = 3.0 x 10$^{-6}$ Ω·m, Q = 0.45Å$^{-1}$, $n_c$ = 1.9 x 10$^{-3}$ Å$^{-3}$, K = 6.2 x 10$^{-3}$ eV·Å$^{-1}$, $r_0$ = 5 x 10$^{15}$ m$^{-1}$s$^{-1}$, $\varepsilon_B$ = 0.024. (The extension to parameters appropriate for blue bronze will be discussed later.)

For each frequency, the strain profile was found for two different boundary conditions. a) "Free contacts": We assumed that the CDW phase was not fixed at the current contacts; this is appropriate for the very short non-perturbative side contacts used by Adelman, *et al*,[3] who showed that in such a case phase slip extends for ~ 100 μm beyond the contacts. For this case, we fixed the phase at zero at a point d ≥ 100 μm beyond the contact; the results did not depend on the choice of d. b) "Pinned contacts": We assumed that the CDW phase was pinned at zero at the current contacts; this would be appropriate for strongly perturbative contacts, such as end contacts or presumably long side contacts, such as those used in our electro-optics experiments.[5,8,9]

For most of our simulations, we took $E_P$ as a constant (= 170 mV/cm, corresponding to a depinning threshold current density $j_T$ = 1.8 x 10$^7$ A/m$^2$). However, Adelman *et al* showed that the pinning field varied with CDW current for their sample, with a minimum value of 170 mV/cm at $j_c = 0$ but increasing rapidly with CDW current, saturating at ~ 350 mV/cm for $j_c > 2\ j_T$.[3] To examine the effects of non-constant $E_P$, we also solved for the strain profiles with this observed $E_P(j_c)$.

The inset to Figure 1 shows the spatial dependence of the strain for pinned contacts with constant $E_P$ for 12.5 kHz and 3 kHz square waves of amplitude $j_{total} = 3\ j_T$ immediately after reversing the current, before the strain has a chance to change. (The spatial dependence is shown for half the sample's length, since the strain is antisymmetric about the center of the sample.) At the lower frequency, this "initial" strain has the expected spatial dependence; i.e. it varies linearly with position near the center of the sample, with extra strain near the contact.[3,5] At the higher frequency, however, the strain stays small for a large region near the center of the sample, as if the strain doesn't have sufficient time to flow from the contacts to the center.[8]

Figure 1a shows the spatial dependence of the strain for a 3 $j_T$, 12.5 kHz square wave at three instants after a current reversal (at t = 0), and Figures 1b,c,d show the time dependences (for a half-period of the square-wave) at three values of x (distance from the contact). For each case, we show the results with pinned contacts and $E_P$=constant (dashed curves), free contacts and $E_P$=constant (solid curve), and free contacts with variable $E_P$ (dotted curves). Note that the magnitude of the strain starts decreasing immediately at the contacts, but much more slowly with increasing x, as also observed by Adelman et al;[3] in fact, away from the contact (Figures 1c, 1d), the magnitude of the strain increases briefly before beginning its reversal. This "delay" time in the strain reversal increases with x.

The effect of variable $E_P$ is to slightly decrease the strain in the phase-slip region, as expected, but does not significantly affect the time evolution of the strain. This was the case for all frequencies and currents examined, so for the remainder of the paper we will discuss only the simpler $E_P$=constant results.

The boundary conditions at the contact cause larger differences. From the spatial dependence, one sees that allowing phase slip beyond the contacts reduces the strain, as expected. This outside phase slip also has the surprising effect of slowing the strain reversal between the contacts, especially in the vicinity of the contacts.

The frequency dependence of the strain, $\varepsilon_\omega$, for a few currents and two positions, is shown in Figure 2. For each case, the filled symbols are for pinned contacts and the open symbols are for free contacts; note that, in each case, the changes in response are slower for free contacts. At x=0, (Figure 2a) the response is essentially relaxational, with a peak in the quadrature response coinciding with the shoulder of the in-phase response. The relaxation time ($1/\omega_{peak}$) decreases with increasing current, as we observed in the electro-optic experiments on blue bronze.[8,9]

Note, however, that the magnitudes of the quadrature peaks are less than half the magnitudes of the in-phase responses at low-frequencies; this is true even at very small currents where phase-slip becomes negligible and the response is essentially diffusive. In the diffusive limit, the relaxation time of the $n^{th}$ *spatial* Fourier component is $\tau_n = (L/n\pi)^2/D$, where the diffusion constant is[9]

$$D = (Q/en_c)^2 K/(\rho_0+\rho_{CDW}). \qquad (5)$$

While the higher spatial components have negligible effect on the quadrature response, they enhance the low-frequency in-phase response (by ~ 20% for pinned contacts).

Away from the contact (Figure 2b), the in-phase response becomes inverted at high frequency, corresponding to the delayed response discussed above.[8] The delay time increases with decreasing current, also as we observed for blue bronze.[8] It is also greater for free contacts than pinned contacts, consistent with the outside phase-slip slowing the overall response.

To parameterize these curves, we fit them to the modified harmonic oscillator expression that we used for the electro-optic response,[8,9]

$$\varepsilon_\omega = \varepsilon_0 / [1-(\omega/\omega_0)^2 +(-i\omega\tau_0)^\gamma], \qquad (6)$$

where $\varepsilon_0$, $\tau_0$, $\omega_0$, and $\gamma$ are current and position dependent. The resonance term corresponds to the delay, with $\tau_{delay} \sim 1/\omega_0$ for values of $\gamma \sim 1$ (as is generally the case for $x \neq 0$, as shown below in Figure 3d). The fits for the pinned contact cases are shown in Figure 2; similar fits are obtained for the free contact cases. The current dependence of the fitting parameters, for both free and pinned contacts, at three positions are shown in Figure 3, while the position dependence of the relaxation and delay times are shown in Figure 4. Figures 3 and 4 again show that the response is faster for pinned contacts than free contacts at every position and current.

Note (Figure 2) that the fits are much better away from the contacts than at x = 0, where the absence of delays (i.e. $\omega_0 = \infty$) reduces the number of fitting parameters. The comparatively poor fits at x=0 reflect the enhancement of the in-phase response with respect to the quadrature compared to perfect relaxation; Eqtn. (6) attempts to accommodate this enhancement by reducing $\gamma$ (for small currents) and decreasing $\tau_0$ by ~ 20% (at all currents) with respect to $1/\omega_{peak}$. Away from the contacts, the expression fits the data very well, except at the highest currents, where it underestimates the magnitude of the inverted in-phase strain at high frequencies.

In general, the relaxation time decreases rapidly with increasing current (Figure 3a), as we also observed in blue bronze.[8,9] The inset to the figure shows the behavior near threshold; as the phase-slip rate vanishes, $\tau_0$ saturates at the diffusion time ($L^2/\pi^2 D$) for pinned contacts. However, for free contacts, $\tau_0$ actually has a maximum at ~ 1.3 $j_T$; this is another unexpected consequence of the outside phase-slip slowing the dynamic response between the contacts. That saturation was observed in our electro-optic experiments on blue bronze[9] is therefore an indication that we had pinned contacts, as expected.

Away from the contacts, the delay times decrease with increasing current (Figure 3b), but more slowly than the relaxation times, so that at the highest currents the response actually becomes underdamped ($\omega_0\tau_0 < 2$). Both time constants increase with distance from the contact. All these features were also experimentally observed in blue bronze.[8] For small currents, the delay time varies linearly with x (Figure 4b); in blue bronze, we interpreted this as flow of strain from the contact with constant velocity,[8] mentioned above. However, this linear dependence does not hold at higher currents.

Also, as we experimentally observed in blue bronze, the amplitude of the response continues growing with current in the phase-slip region, as required for current conversion, but saturates (for $j > 2j_T$) in the "linear" region (Figure 3c),[5,8] where the strain can be viewed as a finite-size effect[4]. In the simulation, however, $\gamma$ stays close to unity for small currents away from the contacts (Figure 3d), whereas for blue bronze we usually observed $\gamma$ to substantially decrease at low currents,[8] which we interpreted in terms of a broadening distribution of relaxation times.

As stated above, these simulation results have several qualitative similarities with our electro-optic results on blue bronze:[8] in particular, the current and position dependences of the amplitude, relaxation time, and delay time. Note, that both time constants in blue bronze are one to two orders of magnitude slower (at T ~ 80 K) than for the simulated results.[8] This is reflected in the different diffusion constants of semiconducting blue bronze and semimetallic NbSe$_3$ (Eqtn. (5)): while blue bronze's CDW elastic constant is an order of magnitude larger than that of NbSe$_3$,[9] its single particle resistivity is ~ 300 times larger (so that $\rho_0 \gg \rho_{CDW}$ for blue bronze).[10] There are important factors which prevent quantitative modeling of our blue bronze results, however:

  a) The blue bronze experiments were voltage driven,[5,8,9] whereas the simulations are current driven. This distinction is not trivial, because the electric field is not constant in the sample and in fact varies with time as well as position,[3] so that a square-wave current does not correspond to a square-wave voltage. This factor prevents quantitative comparisons of measured and simulated current dependences, especially at high frequencies and small currents.[3,11]

  b) Strains in the CDW are screened by quasiparticles, so that the local single particle density, $n_s$, will change with strain, $\Delta n_s = -\Delta n_c = -Q\varepsilon/\pi\Omega$, where $\Omega$ is the area/conducting chain.[1,5] In blue bronze, $n_s$ is activated (e.g. at T ~ 80 K, $n_s \sim n_c/1000$)[10,12] and $\Delta n_s/n_s$ may not be negligible,[13] with the consequence that the single particle resistivity cannot be treated as a constant in Eqtn. (3).[12] However, the CDW elasticity is also approximately inversely proportional to $n_s$,[13] making the diffusion constant (Eqtn. (5)) roughly independent of strain. Our observation that the strains in blue bronze do not pile up on one side of the sample but stay approximately antisymmetric about the center[5,8] therefore suggests that the phase slip rate is independent and pinning field inversely proportional to $n_s$. The latter is expected, because in the three-dimensional "weak" (i.e. collective) pinning limit, $E_P \propto v^4/K^3$, (Ref. [15]) where v is the impurity potential, which will also be screened by quasiparticles and inversely proportional to $n_s$. (Note that these arguments only hold for weak CDW pinning,[1,15] appropriate for the relatively high temperatures at which the electro-optics measurements were done.[5,8,9] At lower temperatures, in the

"strong", hysteretic pinning regime[1,15] $E_P$ was observed to increase with optical excitation of quasiparticles.[16,17])

c) Eqtn. (4) was found to be a good fit to the phase-slip rate by Adelman *et al* for their NbSe$_3$ sample for strains near the contact, but did not fit $r_{ps}$ away from the contact[3] (where the rate of phase-slip is very small and doesn't effect the strain profile significantly). We know of no experimental probes of the strain dependence of phase-slip in blue bronze, however, so it is not clear if a similar expression is valid and, if so, what value of $\varepsilon_B$ is appropriate.

d) In blue bronze, we occasionally observed a decay of the electro-optic signal at long times,[8] but were unable to determine how this decay varied with position or voltage. Such decay is not a feature of Eqtn. (3). It is presumably not a consequence of the electro-optic experiments being voltage-driven, rather than current-driven, since this difference is expected to be most pronounced at high frequencies. We suggest that the decay of the electro-optic response may be a consequence of $E_P$ and $r_{ps}(\varepsilon)$ varying with position, not only distance from the contact but also across the sample cross-section (e.g. enhanced pinning near the surface).

In summary, we have extended the phase-slip model of Reference (3) to study the dependence of CDW strain on the frequency and amplitude of applied square-wave currents. We have found that pinning the phase at the current contacts increases the rate of change of strain everywhere, as compared to allowing the phase to slip outside the contacts, in which case the response actually slows with increasing current near threshold. Adjacent to the contacts, the phase change is essentially relaxational, but a delay develops away from the contacts. Both the relaxation time and delay time increase with increasing distance from a contact and both decrease with increasing current, with the result that the response of the phase becomes underdamped at high currents. All of these effects were observed in electro-optic measurements on blue bronze.[8]

We thank T. Adelman and D. Dominko for helpful comments. This research was supported by the National Science Foundation, Grant # DMR-0400938.

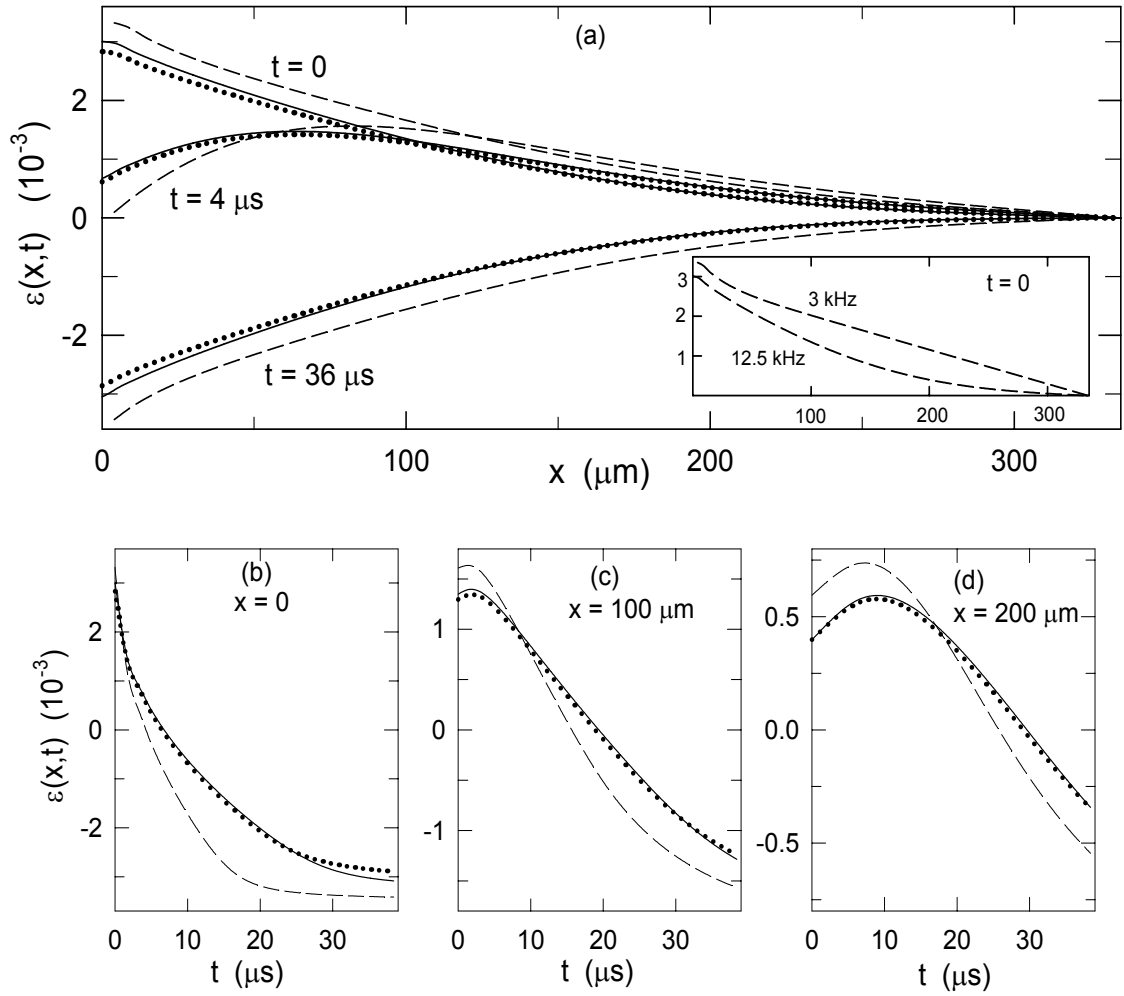

**Figure 1.** Dependence of strain on the distance from a current contact (x) and time after a current reversal for a $j_{total} = 3\, j_T$, 12.5 kHz square-wave. (a) position dependence at three times; (b,c,d) time dependence at three positions. Dashed curves: pinned contacts, $E_P$=constant; solid curves: free contacts, $E_P$=constant; dotted curves: free contacts, $E_P$ variable. Inset: Comparison of initial (t=0) spatial dependences for 12.5 kHz and 3 kHz square waves ($j_{total}=3j_T$, pinned contacts, $E_P$=constant).

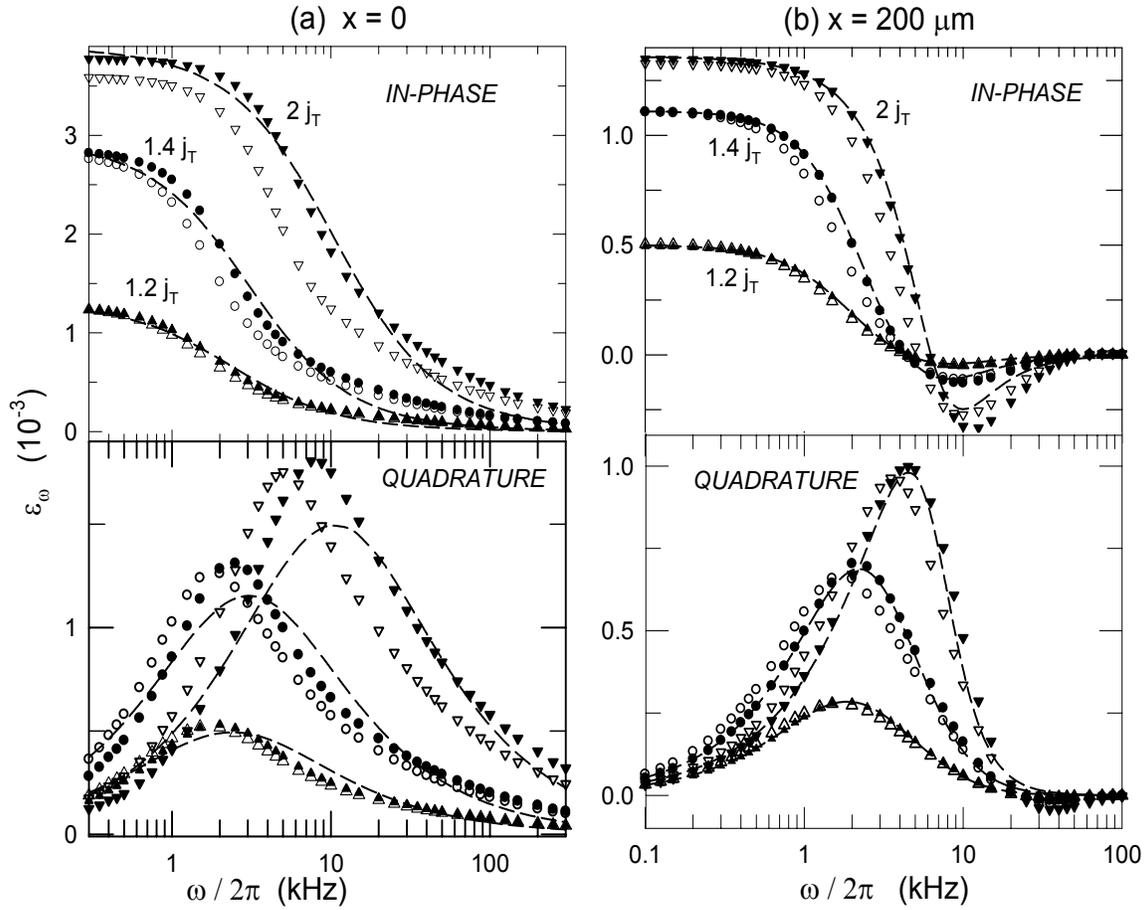

**Figure 2.** Frequency dependence of the strain (a) adjacent to a current contact and (b) 200 μm away, for three values of $j_{total}$. The top panels show the responses in-phase with the applied square-waves and the bottom panels show the quadrature responses. Solid symbols correspond to pinned contacts and open symbols to free contacts (both with $E_P$ = constant). The dashed curves show fits for the pinned contact cases to Eqtn. (6).

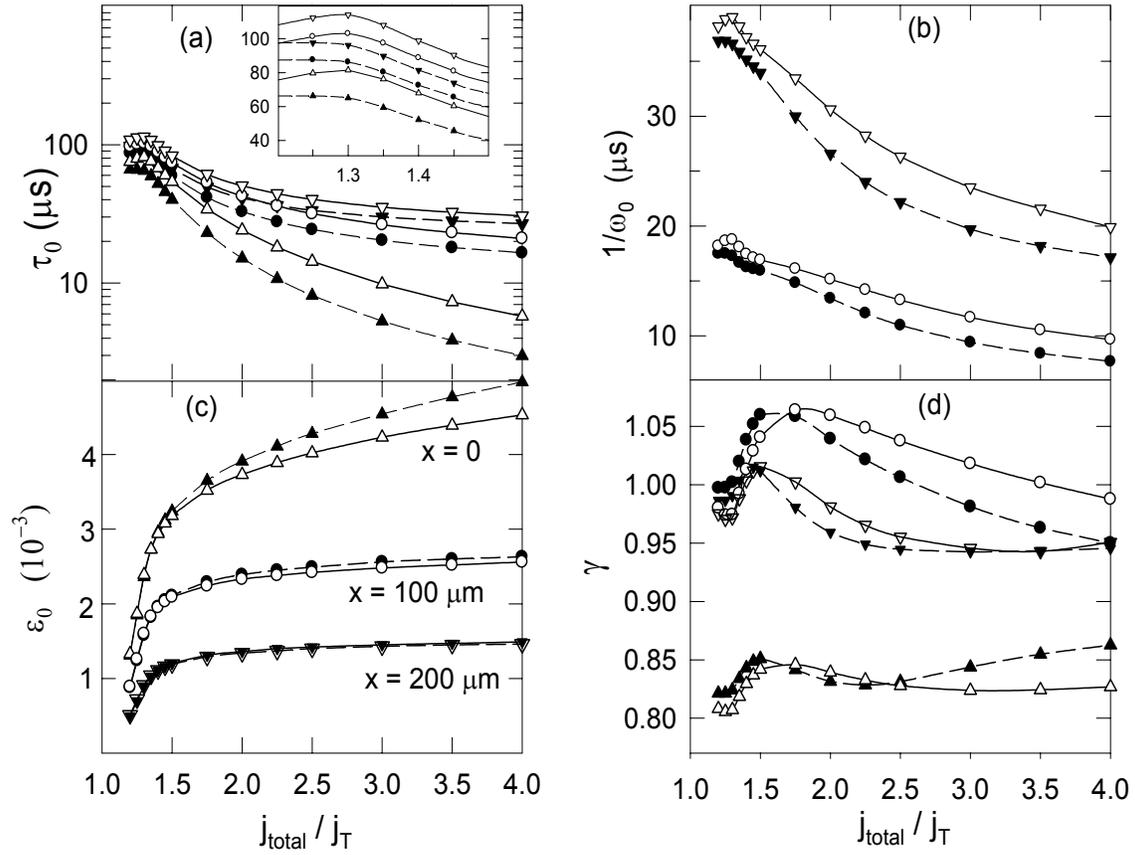

**Figure 3.** Current dependence of the fitting parameters to Eqtn. (6) for x = 0 (up triangles), x = 100 μm (circles), and x = 200 μm (down triangles). Solid symbols (dashed curves) correspond to pinned contacts and open symbols (solid curves) correspond to free contacts, all with $E_P$=constant. (The curves are guides to eye.) Note that at x=0, $1/\omega_0=0$ (not shown). The inset to (a) shows the dependence of $\tau_0$ on small currents in a linear scale.

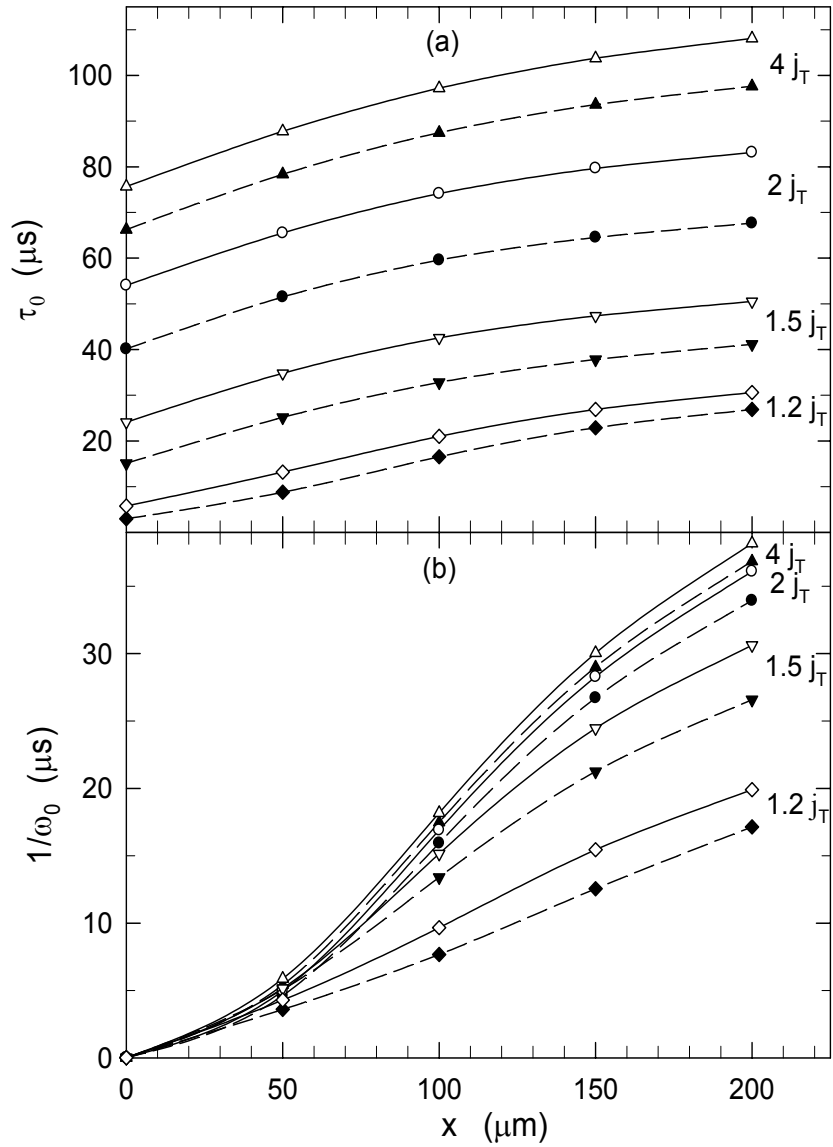

**Figure 4.** Position dependence of (a) average relaxation time and (b) delay time for a few currents. Solid symbols (dashed curves) correspond to pinned contacts and open symbols (solid curves) correspond to free contacts, all with $E_P$=constant. (The curves are guides to eye.)